\documentclass[twocolumn,superscriptaddress]{revtex4-2}

\usepackage{mathrsfs}
\usepackage{amsmath}
\usepackage{amssymb}
\usepackage{graphicx}
\usepackage{float}
\restylefloat{table}

\begin{document}

\title{3D two-electron double quantum dot: comparison between the behavior of some physical quantities under two different confinement potentials in the presence of a magnetic field}

\author{A. M. Maniero}
\email{angelo.maniero@ufob.edu.br}
\affiliation{Centro das Ci\^encias Exatas e das Tecnologias, Universidade Federal do Oeste da Bahia, 47808-021, Barreiras, BA, Brazil}

\author{F. V. Prudente}
\email{prudente@ufba.br}
\affiliation{Instituto de F\'{\i}sica, Universidade Federal da Bahia, Campus Universit\'ario de Ondina, 40170-115, Salvador, BA, Brazil}

\author{C. R. de Carvalho}
\email{crenato@if.ufrj.br}
\affiliation{Instituto de F\'{\i}sica, Universidade Federal do Rio de Janeiro, Rio de Janeiro, 21941-972, RJ, Brazil}

\author{Ginette Jalbert}
\email{ginette@if.ufrj.br}
\affiliation{Instituto de F\'{\i}sica, Universidade Federal do Rio de Janeiro, Rio de Janeiro, 21941-972, RJ, Brazil}

\date{\today }
\begin{abstract}
We have considered a system consisting of two coupled quantum dots containing two electrons, i.e., two quantum dots next to each other with one excess electron each, subjected to an uniform magnetic field perpendicular to the quantum dots plane. The effect of different confinement potential profiles under which the electrons are subjected is studied. This study has been performed in the light of interest in fundamental logical quantum-gate operations: we have been concerned in analysing the behaviour of the main physical quantities which should be involved in a two-qubit quantum-gate operation for two different profiles of confinement found in literature. Our purpose was to establish how sensitive the physical quantities are to the confinement profile and in which range of magnetic field this issue can be critical.
\end{abstract}

\maketitle

\section{Introduction}

In the last few decades, the area of nanostructures has aroused enormous interest due to its impact on computing and information technology as well as on fundamental physics, since nanostructures challenge our knowledge of quantum physics in systems of this scale \cite{kouwenhoven2010spin,jacak1998single,heinzel2008mesoscopic,natelson2015nanostructures}. Nanostructures correspond to confined quantum system whose study of their physical properties can provide a feedback for the design and synthesis processes of new materials, and also new nano and mesoscopic structures. Besides, the study of confined quantum systems covers a broad range of subjects which involve the confinement of atoms or molecules in different environments and conditions \cite{sabin2009v57-58,sen2014electronic,grandjean2018origin}. In this vast area of nanostructures and confined quantum systems one finds the few-electron quantum dots (QDs) \cite{kouwenhoven2001,chan2004} as a particular system; QDs' usefulness ranges from an abstract,  theoretical and fundamental subject such as quantum information \cite{mazurek2014,ayachi2014,roszak2015} to  concrete and experimental applications in areas other than physics or chemistry \cite{efros2018}.

In the line of interest of quantum computation, the study of the  electronic level structure of few-electron QDs as a function of characteristic parameters, which can be experimentally manipulated, is a  relevant issue; the electronic level structure directly affects the exchange coupling $J$, the electronic density function \cite{Boyacioglu_2007,maniero2020a} or, equivalently, the spatial electronic entanglement \cite{khordad2017}. Therefore, by computationally simulating how the electronic levels behave under different experimental conditions (external fields) and different confinement potentials (QDs architecture or their surroundings) is something of practical interest in the search for the basic conditions to be satisfied for realization of fundamental logical operations \cite{burkard1999}.

On the other hand, the computation of macroscopic physical quantities associated to QDs, such as susceptibility and permittivity \cite{mandal2015}, should be able to guide the development in the design of new optoelectronic devices in general \cite{wu2015,yu2020}. The seek of a basis set which allows physical quantities of a macroscopic medium, such as the magnetic susceptibility, to be gauge invariant was an issue addressed by London in the late 30's \cite{london1937} by studying the electronic behavior of a ``macroscopic medium'' -- a chain of atomic centers with one electron each -- in the presence of an uniform magnetic field. This approach proved to be satisfactory \cite{pople1962} and consists of forming a molecular orbital as a linear combination of atomic orbitals in their gauge invariant form
\begin{equation}
\Psi'(\vec r)={\rm exp}[i\Gamma(\vec r)]\Psi(\vec r)
\end{equation}
corresponding to the gauge transformation $\vec A'=\vec A + \nabla\Gamma$, where $\vec A$ is the magnetic vector potential. As noted by Schmelcher, the phase factor associated with the gauge transformation can be seen as a change of origin of the atomic orbital to the location where the atomic center is situated \cite{schmelcher1988}. This specific issue does not matter when one deals with only a single atomic center, even in the presence of a magnetic field, as the atomic center is situated at the origin and, consequently, the complex phase factor can be disregarded \cite{maniero2020a,maniero2021}. Besides, it is also noticed by Schmelcher that, although, London's approach describes diamagnetism properly, it does not guarantee the correct energy expectation value, being, therefore, necessary to minimize with respect to $\Gamma$.

In Ref.~\cite{Boyacioglu_2007}, where two-electron singlet states of a 3D single QD with Gaussian conﬁnement is investigated, it was observed that the values of the physical quantities of practical purpose are pretty close for the gaussian and parabolic potentials. Furthermore, the authors observed that it would be interesting to extend their work to include the effect of an external magnetic ﬁeld $B$ and study the crossing between the singlet and triplet energy levels, because experimentally this would give the signature of the Coulomb correlation effect. Hence, we decided to compare the behaviour of the electronic states of a 3D two-electron double QD  -- two QDs next to each other, with one excess electron each, making them a coupled system -- in the presence of a magnetic field under a gaussian potential, with that of a quartic potential that we had already used earlier~\cite{carvalho2003,maniero2019}. In fact, for a single QD we have observed this crossing between the singlet and triplet energy levels; it is revealed by the oscillatory behavior around the zero value of the exchange coupling $J$~\cite{maniero2020a, maniero2021}. The same does not happen in the case of the double QD, i.e. $J$ does not go back and forth around zero as $B$ strength increases, as we show below; it remains positive for the entire range of $B$ probed.  However, for a certain range of $B$ strength, it presents a back and forth behaviour.

Moreover, Ref.\cite{burkard1999} presents a scenario of how quantum computation may be achieved with a two-electron double QD. That discussion works as further motivation in order to probe the two different confinement potential mentioned above and to analyse the dependence on their form, and consequently to see how accurate the results depend on this aspect; in a real experimental situation the exact form of the confinement potential should be estimated and this can be a limitation in performing a quantum gate operation. Hence in this work we compute the behavior of some physical quantities -- the energy levels, the exchange coupling $J$ and the pair density function -- for the two different profiles of confinement potential, in the presence of a uniform magnetic field, and discuss the outcome obtained.\\ \\

\section{Theoretical Approach}

Our problem is to solve the Sch\"odinger equation $\hat H \Phi = E\Phi$, whose solution $\Phi$ is a linear combination of configuration state functions (CSFs),
\begin{equation}
\Phi=\sum_{l=1}^{N^\textrm{CSF}}C_l^{\textrm{CSF}}\Psi_l^{\textrm{CSF}},
\label{Phi}
\end{equation}
where $N^\textrm{CSF}$ is the number of CSFs, $\Psi_l^{\textrm{CSF}}$ is the $l$th CSF, and $C_l^\textrm{CSF}$ is its corresponding coefficient. Furthermore, a CSF is constituted of linear combination of Slater determinants, i.e.,
\begin{eqnarray}
\Psi_l^{\textrm{CSF}}=\sum_{m=1}^{N^\textrm{det}_l}C_{m}^{\textrm{det}}\textrm{det}_{l,m},
\label{Psi_csf}
\end{eqnarray}
where $N^\textrm{det}_l$ is the number of Slater determinants of the $l$th CSF and   $\textrm{det}_{l,m}$ its $m$th determinant.

The system Hamiltonian reads
\begin{eqnarray}
    \hat{H}=\sum_j^N\hat{O}_1(\vec r_j) +
    \sum_j^N\sum_{n<j}^N\hat{O}_2(\vec r_j,\vec r_n),
\end{eqnarray}
whose operators $\hat{O}_1(\vec r_j)$ and $\hat{O}_2(\vec r_j,\vec r_n)$ are given by
\begin{eqnarray}
    \hat{O}_1(\vec r_j)=\frac{1}{2m_c}\Big[\vec p_j+\vec A(\vec r_j)\Big]^2 + g\mu_B\vec S_j\cdot\vec B_j+\hat{V}(\vec r_j)
\end{eqnarray}
and
\begin{eqnarray}
    \hat{O}_2(\vec r_j,\vec r_n)=\frac{1}{\kappa\vert\vec r_j-\vec r_n\vert},
    \end{eqnarray}
where $g=-0.44$ is the effective giromagnetic factor for GaAs; $\mu_B=\frac{1}{2}$, the Bohr magneton; $\kappa=13.1$, the dielectric constant; and $m_c=0.067$, the effective mass of the electron. We shall use two different types of potentials $(\hat{V}(\vec r_j))$ which we describe later.

As usual, we can take the vector potential $\vec A$ in the Coulomb gauge, 
    \begin{eqnarray}
        \vec\nabla\cdot\vec A=0,
    \end{eqnarray}
    which, for an uniform magnetic field $\vec B$ in the $z$ direction,
\begin{eqnarray}
    \vec B=B\hat{z},
\end{eqnarray}
yields
\begin{eqnarray}
\vec A(\vec r_j)=-\frac{1}{2}\vec r_j\times\vec B=\frac{1}{2}(-y_j\hat{x}+x_j\hat{y})B.    
\end{eqnarray}

Making the same type of considerations employed in the theoretical approach of Ref.\cite{maniero2020a}, in the present case the operator $\hat O_1(\vec r_j)$ can be written as
\begin{eqnarray}
    \hat{O}_1(\vec r_j)&=&  -\frac{\nabla^2_j}{2m_c} + \hat{V}(x_j,y_j,z_j) + \frac{1}{2}m_c\omega_L^2 (x^2_j+y^2_j)   \nonumber \\
    &+&\frac{1}{2}m_c\omega_L^2(x^2_j+y^2_j)+
    \mu_B B\left(\frac{1}{m_c}L_{z_j}+gS_{z_j}\right), \nonumber \\
    &&
\end{eqnarray}
where
\begin{eqnarray}
    \omega_L=\frac{B}{2m_c}
\end{eqnarray}
is the Larmor frequency, and finally the Hamiltonian reads
\begin{eqnarray}
\hat{H}&=& \sum_j^N \left[-\frac{\nabla^2_j}{2m_c}+ \hat{V}(x_j,y_j,z_j) + \frac{1}{2}m_c \omega_L^2(x^2_j+y^2_j) \right.\nonumber \\
&+&\left. \frac{\mu_B B}{m_c}L_{j_z}+\mu_B B g S_{z_j}\right] + \sum_j^N\sum_{n<j}^N\frac{1}{\kappa\vert r_j-\vec r_n\vert}.
\end{eqnarray}

\begin{figure}[t]
\vspace*{-0.5 cm}
\hspace*{-0.5 cm}
\includegraphics[scale=0.55]{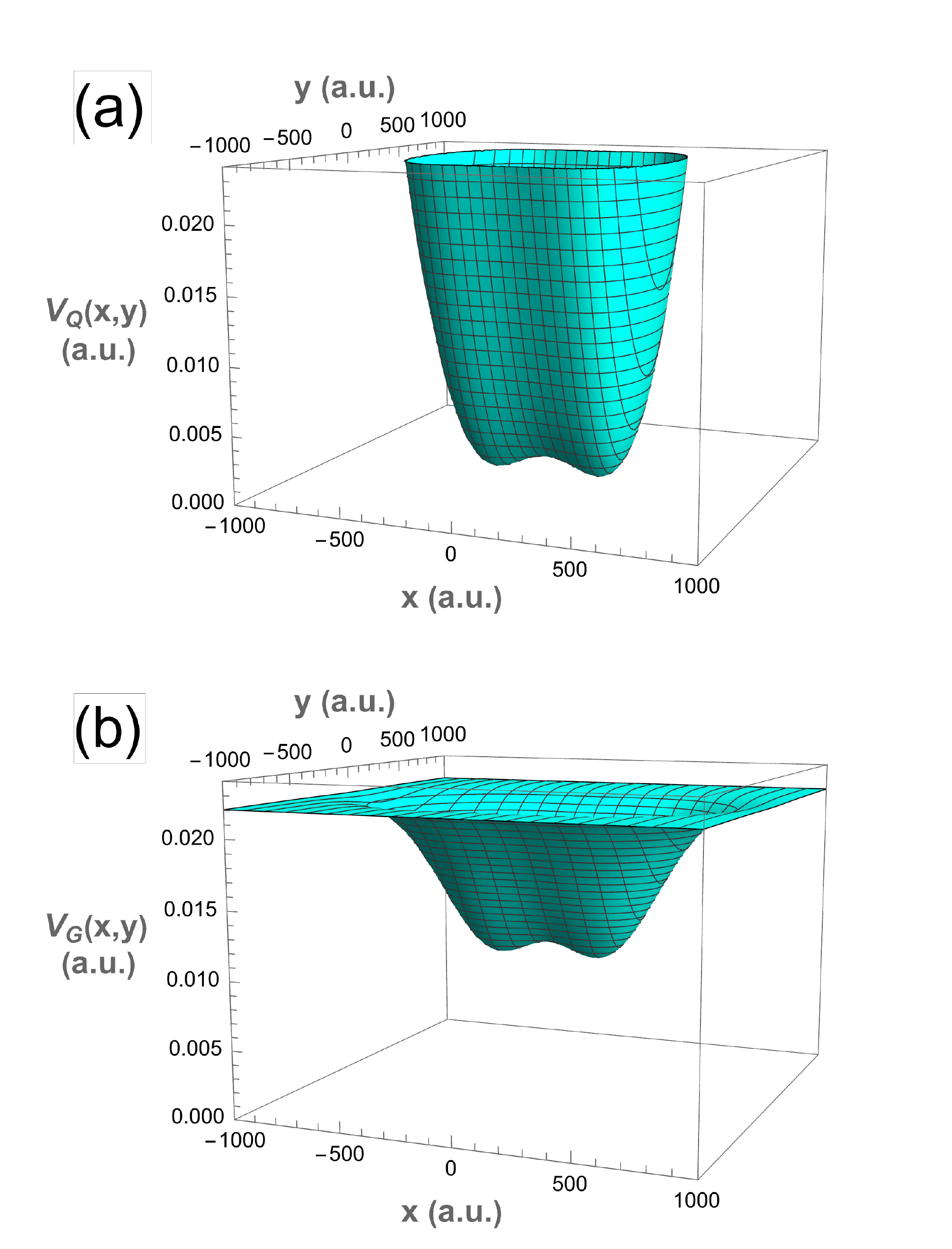}
\caption{3D vision of the confinement potentials: (a) the quartic potential $V_Q(x,y) = V_{Qx}(x)+V_y(y)$ with $X_Q=242.55$, $m_c=0.067$, and $\omega_x=\omega_y=0.001528$; (b) the inverted gaussian potential ${V}_{Gxy}(x,y)$ with $V_0 = 0.0110248$, $\beta = \frac{1}{2 R^2}$, $R=226.767$ and $X_G= 270$. In the subsection \ref{Results-set-param-potentials} we discuss our choice of these parameters' values. }
\label{VQ+VG_3D}
\end{figure}

As mentioned above, we employ two types of confinement potential: a quartic one, which we have already used in previous works, \cite{carvalho2003,maniero2019}
\begin{eqnarray}
    \hat{V}_Q(x_j,y_j,z_j) &=& V_{Qx}(x_j)+V_y(y_j)+V_z(z_j) \nonumber \\
    &=& \frac{1}{8 X_Q^2}m_c\omega_x^2(x^2_j-X_Q^2)^2 + 
    \frac{1}{2}m_c\omega_y^2 y^2_j  \nonumber \\
    &+& \frac{1}{2}m_c\omega_z^2z^2_j;
    \label{quartico}
\end{eqnarray}
and one with a gaussian profile (gaussian in the $XY$ plane and parabolic in the $z$ direction), 
\begin{eqnarray}
\hat{V}_G(x_j,y_j,z_j) &=& V_{Gxy}(x_j,y_j) + V_z(z_j) \nonumber \\
&=& V_0\Big[2 - e^{-\beta(x_j-X_G)^2-\beta y^2_j} \nonumber \\
 &&\;\;\;\;\;\;\; - e^{-\beta(x_j+X_G)^2-\beta y^2_j}\Big]  \nonumber \\ 
&&+\frac{1}{2}m_c\omega_z^2z^2_j.
\label{gausinv}
\end{eqnarray}
One finds in literature the use of gaussian profile to describe the confinement in a two-electron single QD \cite{Adamowski2000,Xie2003,Boyacioglu_2007,chaudhuri2021} as well as in a two-electron double QD \cite{dybalski2005}. In this last quote the confinement potential of the the double QD is achieved by using the superposition of a parabolic profile with a gaussian one; the latter working as the barrier between the two dots. We have made a different option to describe a two-electron double QD; we have opted to choose the confinement potential in the plane $XY$ as a superposition of two inverted gaussians, in order to allow a continuum of states, which leads to the possibility of free electron states in the $XY$ plane. This can be seen in Fig.\ref{VQ+VG_3D}, where it is displayed in panel (a) the quartic potential $V_{Qx}(x)+V_y(y)$, which is limited in the vertical direction only below, whereas the inverted gaussian potential $V_{Gxy}(x,y)$, displayed in panel (b), is delimited both above and below. Thus, the former allows only bound states while the latter allows unbound states which can represent electric currents into and from the double QD. Observe that the lower limit of the quartic potential is at the zero energy level, while for the gaussian one it is higher.

\subsection{Setting the bases}

We have used a computational code to study the energy spectrum of two electrons confined in a three-dimensional (3D) anisotropic potential. In order to employ it, we have had to establish anisotropic orbitals as the atomic basis set. Similar to what was done in our previous works~\cite{olavo2016,maniero2019,maniero2020a}, we have chosen a basis set composed of the Cartesian anisotropic Gaussian-type orbitals (c-aniGTO) centered in the position $\vec R=(X_C,Y_C, Z_C)$ which, apart a normalization constant, are given by
\begin{eqnarray}
g_\mu(\vec r-\vec R ,\vec \alpha)=
(x-X_C )^{n_x}
(y-Y_C )^{n_y}
(z-Z_C )^{n_z}\times \nonumber\\
\exp[ -\alpha_x(x-X_C )^2 - \alpha_y(y-Y_C )^2 - \alpha_z(z-Z_C )^2 ],~
\label{gaussianxyz}
\end{eqnarray}
where one has the possibility of providing different exponents $\alpha_x$, $\alpha_y$ and $\alpha_z$ according to the problem analysed and $\mu$ stands for $(n_x,n_y,n_z)$. In addition, in analogy to the standard convention for the atomic case, we shall  classify the orbitals as $s$-, $p$-, $d$-,... type according to $n=n_x + n_y + n_z = 0, 1, 2,...$, respectively.

The choice of the exponents  $\alpha_x$, $\alpha_y$ and $\alpha_z$ is discussed in the next two subsections. We wish to draw attention to an issue involving the choice of the exponents. As mentioned in the Introduction, the presence of a non-zero magnetic field introduces a phase factor in the basis functions which is related to a gauge transformation. In principle, this forces one to deal with complex basis functions. Although in our previous works we have dealt with real basis functions, since we had $\vec B=0$, and have adopted a certain procedure to obtain the exponents, here we  repeat the same procedure for both zero and non-zero magnetic field regime. However, for the non-zero magnetic regime we have used an adaptation in the procedure that involves identifying the harmonic oscillator frequency $\omega_i$ corresponding to the exponent $\alpha_i$ ($i$ stands for $x,y,z$) and composing it with the Larmor frequency $\omega_L$ associated with the non-zero magnetic field. Finally, it is worth mentioning that when one deals with only a single atomic center, even in the presence of a magnetic field ($\vec B\ne 0$), the complex phase factor in the basis functions can be disregarded and, consequently, one can use real basis function; this is the case of Ref.~\cite{maniero2020a}.

\subsubsection{Without Magnetic Field}

First we have considered the quartic potential (see Eq.(\ref{quartico})). Since the potential $\hat{V}_Q(x,y,z)$ has the form $\hat{V}_Q(x,y,z)=V_{Qx}(x)+V_{y}(y)+V_{z}(z)$ and along the $y$ and $z$ directions it has the same form of the potential used in  work~\cite{olavo2016}, i.e. a parabolic one $V_q(q)=\frac{1}{2}m_c\omega_q^2q^2$ ($q$ stands for $y$ and $z$), the same two types of exponents have been considered here in these directions: 
\begin{equation}
\alpha_i^{(1)}=\frac{m_c \omega_i}{2} \mbox{ and } \alpha_i^{(2)} = \frac{3}{2}\alpha_i^{(1)},
\label{alfa_i}
\end{equation}
where $i$ stands for $y$ and $z$. 

On the other hand, we have verified that the first type of exponent in the $x$ direction, $\alpha_x^{(1)}$, has produced better results when we have used a variational method minimizing the following functional,
\begin{eqnarray}
E(\alpha) = \frac{\int dx \Psi(\alpha,x)\hat{H}\Psi(\alpha,x)}{\int dx  \Psi(\alpha,x)\Psi(\alpha,x)}
\end{eqnarray}
where $\hat H(x) = \left[-\frac{1}{2m_c}\frac{d^2}{dx^2}+\frac{m^*_c\omega_x^2}{8a^2}(x^2-X_Q^2)^2\right]$. Once the potential displayed in the operator $\hat H(x)$ has minima in $x = \pm X_Q$, the functions  $\Psi(\alpha,x)$ have been taken as linear  combination of the functions centered at the same points. This means that they correspond to $x$ direction molecular orbitals given by
\begin{eqnarray}
\Psi(\alpha,x) = e^{-\alpha(x-X_Q)^2} +e^{-\alpha(x+X_Q)^2}
\end{eqnarray}
Finally, the second type of the exponent has been chosen similarly as the second type of the $y$ and $z$ exponents, namely $\alpha_x^{(2)} = 3\alpha_x^{(1)}/2$.

Now, considering the gaussian potential, Eq.(\ref{gausinv}), one observes that this potential is not  separable, since its form is

\begin{eqnarray}
\hat{V}_G(x,y,z) = V_{Gxy}(x,y) + V_z(z),
\end{eqnarray}
where
\begin{eqnarray}
V_{Gxy}(x,y)&=&V_0[2-e^{-\beta(x-X_G)^2-\beta y^2} \nonumber \\
&&-e^{-\beta(x+X_G)^2-\beta y^2}]
\label{VG(x,y)}
\end{eqnarray}
and $V_z(z)$ has the parabolic profile $\frac{1}{2}m_c\omega_z^2z^2$. Therefore, as done in the previous case, in the $z$ direction we have considered 
\begin{equation}
\alpha_z^{(1)}=\frac{m_c \omega_z}{2} \mbox{ and } \alpha_z^{(2)} = \frac{3}{2}\alpha_z^{(1)},
\label{alfa_i}
\end{equation}
whereas for the $x$ and $y$ directions the first type of the exponent $\alpha$, i.e. $\alpha^{(1)}$, has been obtained by minimizing 

\begin{eqnarray}
E(\alpha) = \dfrac{\int dx dy \Psi(\alpha,x,y)\hat{h}\Psi(\alpha,x,y)}{\int dx dy  \Psi(\alpha,x,y)\Psi(\alpha,x,y)},
\end{eqnarray}
with respect to $\alpha$, where
\begin{eqnarray}
\Psi(\alpha,x,y) = e^{-\alpha(x-X_G)^2-\alpha y^2} +e^{-\alpha(x+X_G)^2-\alpha y^2}
\end{eqnarray}
and
\begin{eqnarray}
\hat{h}=-\frac{1}{2m_c}\nabla^2+{V}_{Gxy}.
\end{eqnarray}
Hence, $\alpha^{(1)}$ has been obtained by satisfying the condition
\begin{eqnarray}
\frac{d}{d\alpha^{(1)}}E(\alpha^{(1)})=0
\label{minimaenergias}
\end{eqnarray}
which corresponds to the minimum of energy, while the second exponents follows the criterion adopted in the other cases
\begin{eqnarray}
\alpha^{(2)}=\frac{3}{2}\alpha^{(1)}.
\end{eqnarray}

Now, considering the excitations levels given by $(n_x,n_y,n_z)$, as we are interested in confining only along the $z$ direction, we shall use larger values for the $\omega_z$. Consequently we expect few excitation in this direction, i.e., we shall take only $n_z = 0, 1$, whereas in the plane $xy$ we will consider larger values: $n_x, n_y = 0, 1, 2,...$

\subsubsection{Basis with magnetic field}

 In order to take into account the effect of the magnetic field we proceeded as follow. Consider the basis function centered on the origin
\begin{eqnarray}
	\xi_\mu(\zeta,\vec r)= 
	x^{n_x} y^{n_y} z^{n_z} \exp\Big(-\zeta_xx^2- \zeta_yy^2- \zeta_zz^2\Big).
\end{eqnarray}

In respect to it and following the approach adopted by London \cite{london1937}, we can use the  above basis function to construct the set of basis functions for our system, consisted of two centers localized in $(\pm X_C,0,0)$ in the presence of a uniform magnetic field $\vec B=B\hat z$, according to:
\begin{eqnarray}
	\xi_\mu(\zeta,\vec r-\vec R_C)&=& 
	(x \pm X_C)^{n_x} y^{n_y} z^{n_z} \times \nonumber \\
	&& \exp\Big(\pm\frac{iX_C\,y}{2\ell_B\,^2}\Big) \times \nonumber \\
	&& \exp\Big[-\zeta_x(x \pm X_C)^2- \zeta_yy^2- \zeta_zz^2\Big],~~~~~
	\label{f(r)_base_B.NE.0}
\end{eqnarray}
where the displacement from the origin, $\vec R_C=(\pm X_{_C},0,0)$, corresponds to the gauge transformation $\vec A=B(-y,x,0)/2 \rightarrow \vec A=B(-y,x \pm X_{_C},0)/2$ which introduces the phase shift $\exp(\pm iX_C\,y/2{\ell_B}^2)$; besides, $\vec R_C=(\pm X_{_C},0,0)$ is the centers positions of our confinement potential along the $x-$axis and $\ell_B$ is the {\it magnetic length} defined as $\ell_B=\sqrt{\dfrac{\hbar}{eB}}$.
It is convenient to define $k_B=\dfrac{1}{\ell_B}=\sqrt{\dfrac{eB}{\hbar}}$
and  $ Y_C^\prime=\dfrac{X_Ck_B^2}{4\zeta_y}$
so that Eq.(\ref{f(r)_base_B.NE.0}) can be rewritten in the form
\begin{eqnarray}
	\xi_\mu(\zeta,\vec r-\vec R_C) &=&
	(x\pm X_C)^{n_x} y^{n_y} z^{n_z} \exp \Big( -\zeta_y{Y_C^\prime}^2 \Big) \times \nonumber \\
	&&\exp \Bigg[ -\zeta_x\Big(x\pm X_C\Big)^2  \nonumber \\
	&&\quad\quad\; - \; \zeta_y\Big(y\mp iY_C^\prime\Big)^2- \zeta_zz^2\Bigg].
\end{eqnarray}

Now, in the same fashion we have done in previous works, the presence of the magnetic fields changes the exponents in the $x$ and $y$ directions according to
\begin{eqnarray}
\zeta_{1_x} = \frac{m_c}{2}\sqrt{\omega_x^2+\omega_L^2}, &&
\zeta_{1_y} = \frac{m_c}{2}\sqrt{\omega_y^2+\omega_L^2} \nonumber \\
&\mbox{and}& \nonumber \\
\zeta_{1_z} &=& \frac{m_c}{2}\omega_z.
\end{eqnarray}
Consequently, the second type of exponents yield
\begin{eqnarray}
\zeta_{2_x} = \frac{3m_c}{4}\sqrt{\omega_x^2+\omega_L^2}, &&
\zeta_{2_y} = \frac{3m_c}{4}\sqrt{\omega_y^2+\omega_L^2} \nonumber \\
&\mbox{and}& \nonumber \\
\zeta_{2_z} &=& \frac{3m_c}{4}\omega_z,
\end{eqnarray}
where $\omega_x = \dfrac{2\alpha_x}{m_c}$ and $\alpha_x$ is the exponent obtained with the minimizing process of the previous section, without magnetic field.

 All the results displayed in this work were obtained with a basis of  80 functions  (2s2p2d2f) in each  center, with  3,240 (3,160) CSF's and  6,400 (3,160) determinants for the singlet (triplet) states.

\subsection{Pair Density function}
\label{I(x)}
It would be interesting to determine the pair density function and study how it varies with the strength of the magnetic field and the two different types of confining potentials. It is defined as \cite{Boyacioglu_2007} 
\begin{eqnarray}
I(\vec r) =  \int d\vec r_1 d\omega_1 d\vec r_2 d\omega_2 \Phi^\ast  \delta^{(3)}(\vec r-\vec r_1+\vec r_2)  \Phi 
\label{I(r)}
\end{eqnarray}
where
\begin{eqnarray*}
\delta^{(3)}(\vec r -\vec r_1+\vec r_2) &=& \delta(x -x_1+x_2) \delta(y -y_1+y_2) \times  \\
&&\quad \delta(z -z_1+z_2),
\end{eqnarray*}
and $\vec r_j$ is the position of the $j^{th}$ electron ($j=1,2$).

The pair density function is related with the issue of the double-occupation of a single dot by the two electrons, which was addressed in Ref.\cite{carvalho2003}. Reminding briefly, it is known that two-qubit quantum-gate operation involves two coupled quantum dots labeled 1 and 2, each containing one excess electron  with spin 1/2; the dots furnish the tag for each qubit, which is related with the spin of each dot-trapped electron; the two-qubit operation rotates one of the tag-spin, whereas the other one is held in a well deﬁned unchanged direction, via exchange interaction $J$ between the electrons in each dot. Therefore is highly undesirable that the two electrons jump onto the same single quantum dot, because tag information is lost. Consequently, in terms of $I(\vec r)$, it is undesirable to have $I(\vec r=0) \ne 0$.

Finally, from the definition of $I(\vec r)$, Eq.(\ref{I(r)}), and remembering that $\Phi$ is obtained within the Full Correlation Interaction (FCI) approach, expressed by the Eqs.(\ref{Phi}) and (\ref{Psi_csf}), it is easy to show that 
\begin{eqnarray}
I(\vec r) =  I(x) I(y) I(z).
\label{IxIyIz}
\end{eqnarray}

\section{Analysis and Discussion}
\subsection{Setting the Potential Parameters}
\label{Results-set-param-potentials}
Next, we discuss the choice of the parameter values for both confinement potentials. Our choice was oriented to obtain a good agreement between the profiles of the two potentials in the widest possible range along the $x$ direction in the barrier region.

With regard to the gaussian potential, Eq.(\ref{gausinv}), we have taken as reference for the potential parameters those employed in Ref.~\cite{Adamowski2000}, which has been used elsewhere in the literature \cite{Boyacioglu_2007,Xie2003}. Therefore, assuming the double QD as a GaAs/AlGaAs heterostructure, one has the donor Rydberg $R_D = 6$ meV ($\approx 2.20496\times 10^{-4}$~a.u.) and the donor Bohr radius $a_D = 10$~mm ($\approx 188,973$~a.u.). Thus, in accordance with \cite{Adamowski2000} the depth of the potential well have been taken as $V_0 = 50R_D = 300$~meV ($\approx 1.10248\times 10^{-2}$~a.u.). Naming $R$ as the width of the potential well, one has $\beta = \frac{1}{2 R^2}$; we have chosen $R=12$~nm ($\approx 226.767$~a.u.). Besides, we have set $X_G \approx 14.2878$~nm ($270$~a.u.) and $\hbar\omega_z \approx 27.2114$~eV ($1$~a.u.).

\begin{figure}[H]
\hspace*{-1cm}
\includegraphics[scale=0.365]{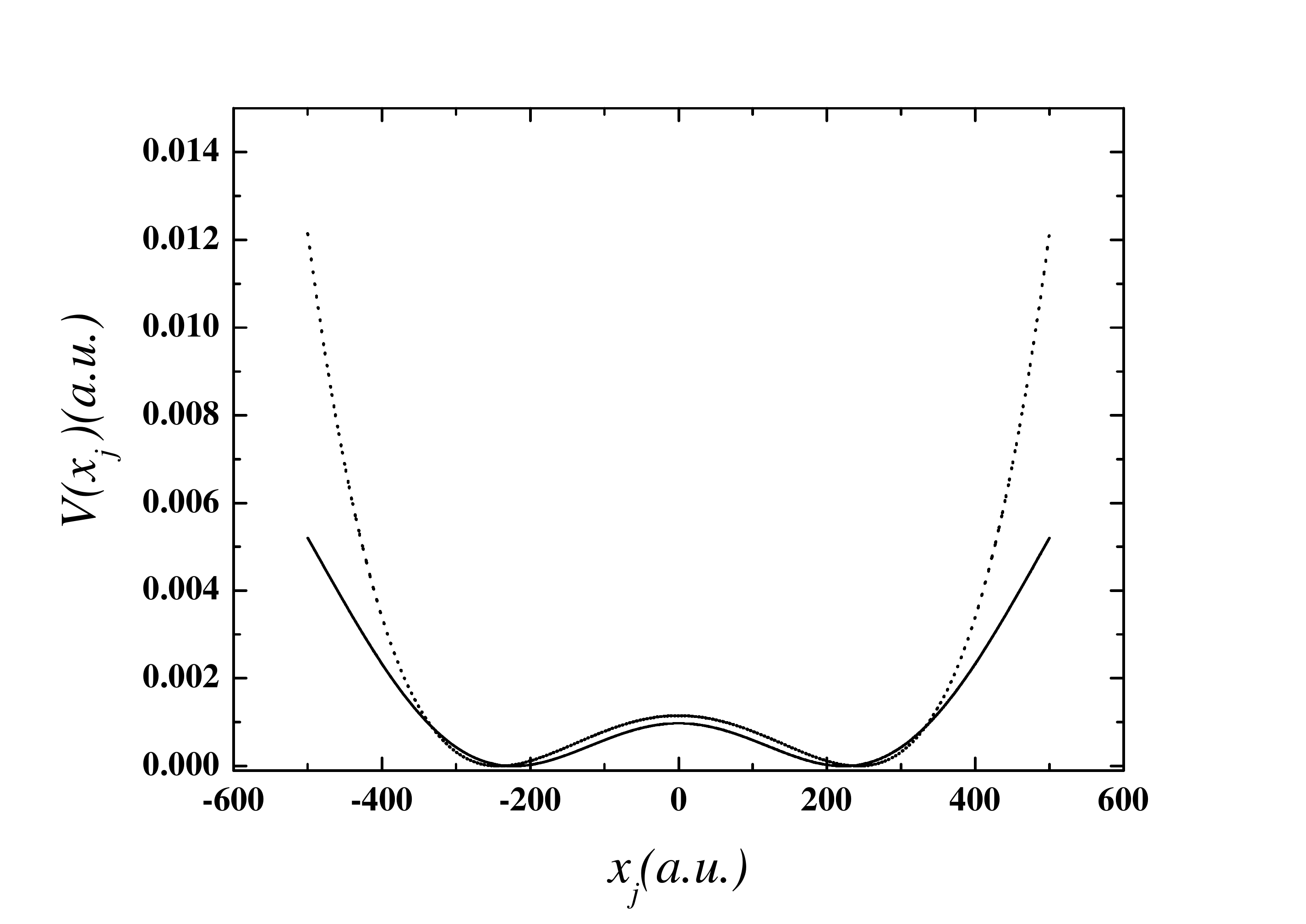}
\caption{Profiles of the quartic and gaussian confinement potentials along the $x$ direction: $V(x_j)=V_{Qx}(x_j)$ (dotted line), and $V(x_j)={V}_{Gxy}(x_j,0)$ (solid line), respectively, where $j$ stands for the label of the $j^{\textrm{th}}$-electron. The graphic parameters are the following. For $V_{Qx}(x_j)$: $X_Q=242.55$, $m_c=0.067$, and $\omega_x = 0.001528$. For ${V}_{Gxy}(x_j,0)$: $V_0 = 0.0110248$, $R=226.767$ ( $\beta = 1/(2 R^2)$), and $X_G= 270$. In fact, in the case of the gaussian potential, the curve plotted is  ${V'}_{Gxy}(x_j,0)={V}_{Gxy}(x_j,0)+C$, where $C=-1.02214\times 10^{-2}$.}
\label{GI-Q-fundo}
\end{figure}

In order to compare the results of the two types of confinement potentials we have adjusted their bottom as is depicted in Fig.\ref{GI-Q-fundo}, adjusting the profile of the quartic potential to the gaussian one. Therefore, for $X_G=270$ we have got $X_Q=242.55$, which is the real place of the bottom of both potentials; as $V_{Gxy}(x,y)$ consists of the addition of two inverted gaussian functions, its two minima will not coincide with the minimum of each inverted gaussian, i.e. $\pm X_G$, they are slightly shifted towards the origin. We have also obtained $\omega_x=\omega_y = 1.52800\times 10^{-3}$, whereas the value $\omega_z=1$ is kept fixed. Besides, according to Eq.(\ref{gausinv}) and as shown in Fig.~\ref{VQ+VG_3D}, the bottom of the  gaussian potential ${V}_{Gxy}(x,y)$ is not at the zero energy level, it is higher. Thus, to make the adjust as displayed in Fig.\ref{GI-Q-fundo} it was necessary to add the value $C=-1.02214\times 10^{-2}$ to it. 

\begin{figure}
\hspace*{-0.8 cm}
\includegraphics[scale=0.365]{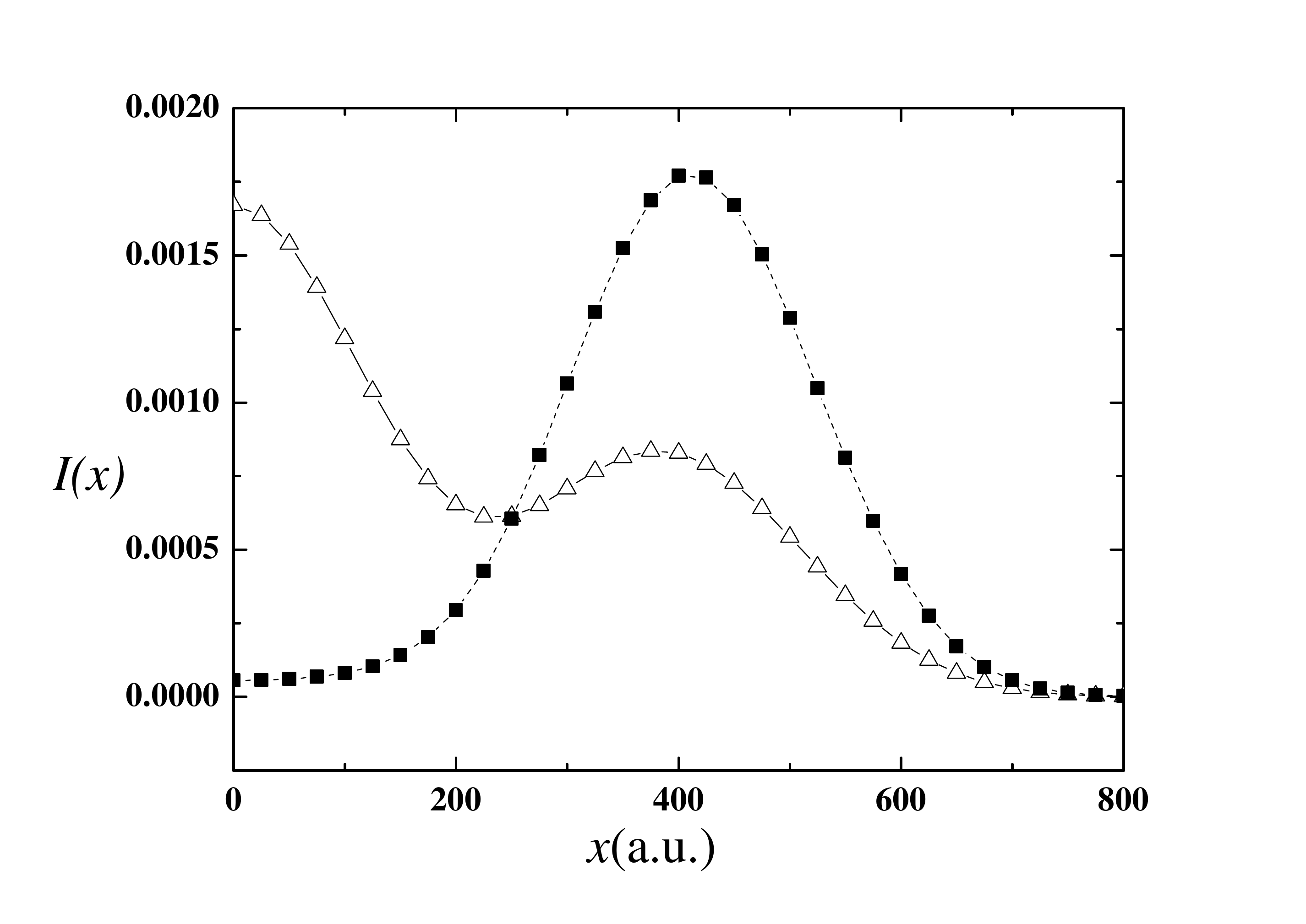}
\caption{$I(x)$ as function of $x$, where $x=x_1-x_2$ (a.u.), in the case of the gaussian confinement ($V_{Gxy}(x_j,0)$) for the singlet state with $B=0$: HF (open triangle black) and FCI (full square black)  computation.}
\label{HF-FCI}
\end{figure}

Before going further in the discussion of the effect of the magnetic field on the electronic structure and the comparison between the two types of potentials, we show in Fig.~\ref{HF-FCI} the results for the density $I(x)$ obtained in the HF and FCI calculation.
The effect of electronic correlation within the FCI method on the computed results is quite clear: in the FCI calculation the pair density is almost zero at the origin ($x = 0$) and has only a single maximum along $x$, which is expected, while in the HF calculation the result points to an absolute maximum of the pair density at $x = 0$ and a secondary one almost at the same position of the FCI's maximum. Due to this clear failure in the HF calculation, we shall only present from now on results obtained with FCI calculations.

\subsection{Numerical Results}

\begin{figure}
\hspace*{-1.2 cm}
\includegraphics[scale=0.6]{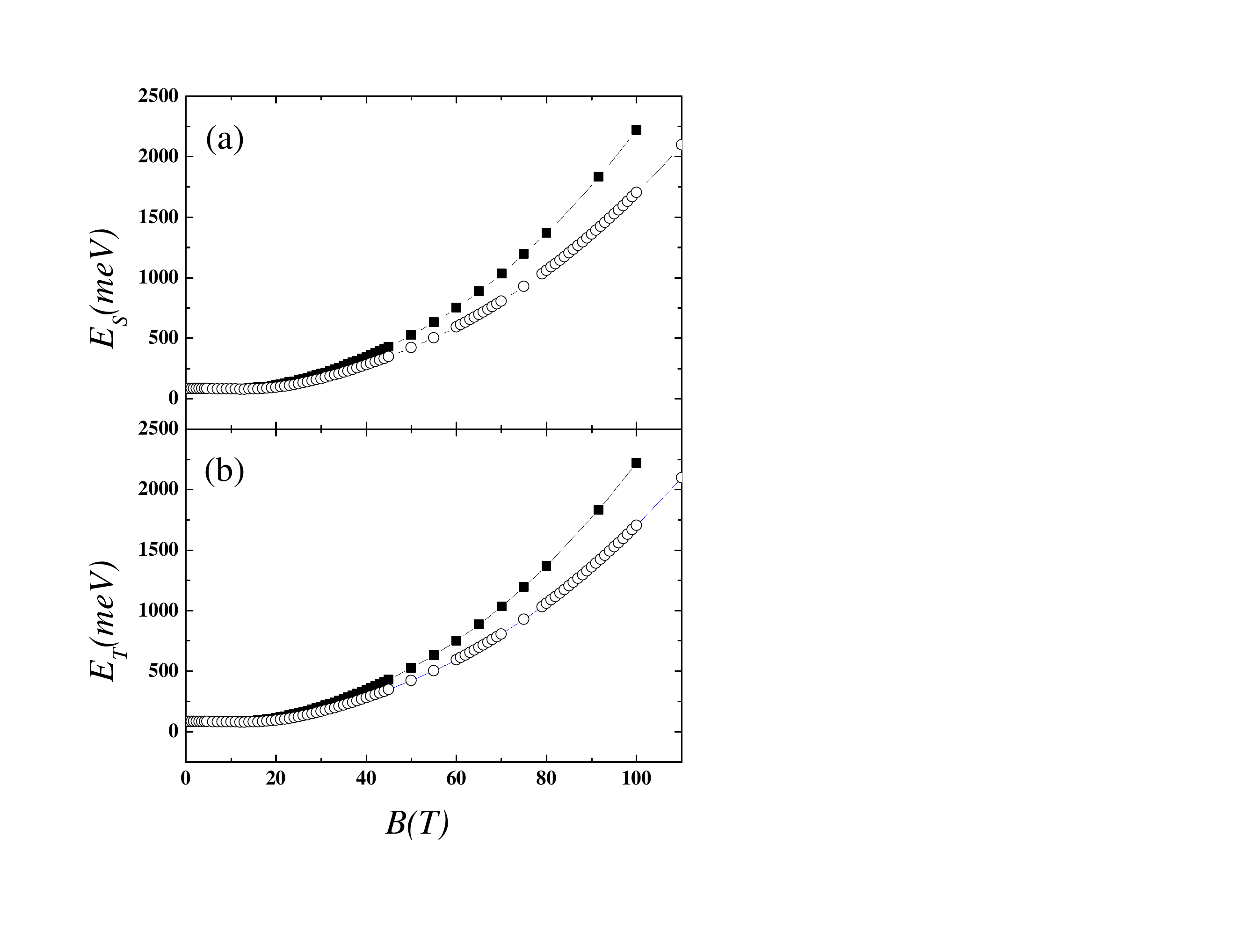}
\vspace{-1.2 cm}
\caption{(a) singlet and (b) triplet state energies as a function of the magnetic field $B(T)$ computed for the gaussian (solid squares) and  quartic  potentials (open circles).}
\label{Energia STXB}
\end{figure}

In Fig.~\ref{Energia STXB} we display the behaviour of the energy of the lowest singlet (a) and triplet (b) states computed for both confinement potentials, the gaussian and the quartic one, as a function of the magnetic field. It is worthy mentioning that to compare these energies we have subtracted the value of 1 a.u. of the energy, which is common in the two cases of confinement and corresponds to the harmonic potential in the $z$-direction acting on the two electrons. Besides, we have also added $2C=2.04428\times 10^{-2}$ a.u. to the energies calculated for the quartic potential; this value corresponds to the difference between the bottom of the two potentials, see Fig.~\ref{VQ+VG_3D}, applied for the two electrons. From now on, we express energy and magnetic field in terms of meV and tesla, respectively; only length remains in atomic units (a.u.). Hence, observing Fig.~\ref{Energia STXB}(a) one can notice that for both potentials the values obtained for the energies of the singlet state are quite close up to $B \approx 25$~T, moving away from there. The same behaviour happens to the energies of the triplet state and is displayed in Fig.~\ref{Energia STXB}(b).

In addition, Fig.~\ref{Energia STXB} reveals an interesting issue concerning the confinement of the two electrons. Regarding the gaussian potential, as the energy difference between the top and bottom of the $V_{Gxy}(x,y)$ is approximately $321.85$~meV, the two electrons are certainly no longer bounded, or confined, due to the gaussian confinement potential when the system reaches energy values greater than $643.7$~meV. From this point on, the energy values as well as their quantization are due only to the confinement provided by $B$. According to Fig.~\ref{Energia STXB} this condition happens for $B \approx 55$~T for both $E_S(B)$ and $E_T(B)$. This confinement effect due to $B$ still remains when electrons are subjected to the quartic confinement potential $V_Q$. In fact, in this case, the electrons are subjected to two confinement mechanisms whatever the energy of the system, since $V_Q$ is not bounded from above. Thus, it would be expected that under a regime of greater confinement -- $B$ plus $V_Q$ -- the energy values as a function of $B$ would be higher than in the case of the gaussian potential, in particular where only $B$ provides the confinement. Nevertheless, Fig.~\ref{Energia STXB} shows us exactly the opposite. This is a question to be analysed in the future: to understand what physics is behind this.

As a general fact, despite the issue raised in the previous paragraph and the discrepancy between the energy values regarding the two  confinement potentials from $B \approx 25$~T onward, what Fig.~\ref{Energia STXB} shows is that the energy curve profiles for the singlet and triplet states, $E_S(B)$ and $E_T(B)$, respectively, are basically the same for any value of $B$ regardless of the confinement: they are smooth and monotonically increasing functions of $B$. 

\begin{figure}[t]
\hspace*{-0.75 cm}
\includegraphics[scale=0.37]{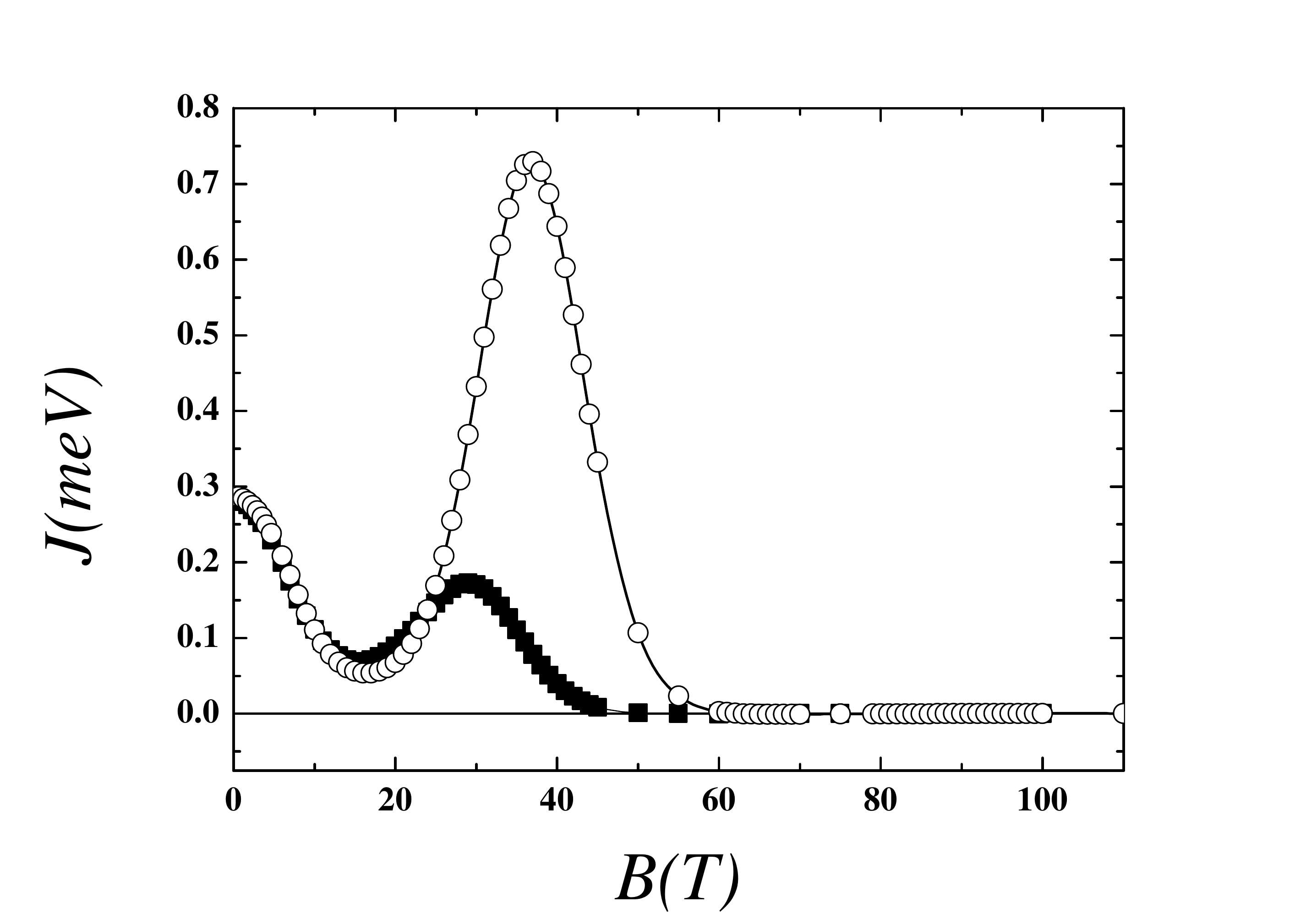}
\caption{Exchange coupling $J$ (a.u.) as a function of the magnetic field $B$ (T) under the two confinement potentials: gaussian potential (solid square) and quartic potential (open circle)}
\label{J versus B}
\end{figure}

Now, let us return to a point addressed by Burkard, Loss and di Vincenzo \cite{burkard1999}. As mentioned previously the exchange interaction $J$ plays an important role in two-qubit quantum-gate operation. Thus, its handling and control is critical and, consequently, the accuracy of its measurement. In Fig.~\ref{J versus B} we display the behaviour of the difference between the energy of the lowest triplet and singlet states, i.e. $J=E_T - E_S$, as a function of $B$ for both type of confinement we are dealing with. We observe that up to $B \approx 25$~T the results provided by the two confinement potentials are basically the same; the plot of $J(B)$ has approximately the same profile and values. However, in the range $25\textrm{~T} \lesssim B \lesssim 60\textrm{~T}$, although the profile of $J(B)$ is similar for both potentials, the actual $J(B)$ values predicted by the two potentials are completely different. What this result shows is that $J(B)$ has a great sensitivity to the confinement profile on this interval of $B$ and, since in an experimental set-up the information available on the confinement profile is poor, one should avoid working in this interval of $B$; one can estimate a value of $J(B)$ completely wrong and then the computation based on $J$ is compromised. \\

\begin{figure}[t]
\vspace*{-0.6 cm}
\hspace*{-1.2 cm}
\includegraphics[scale=0.37]{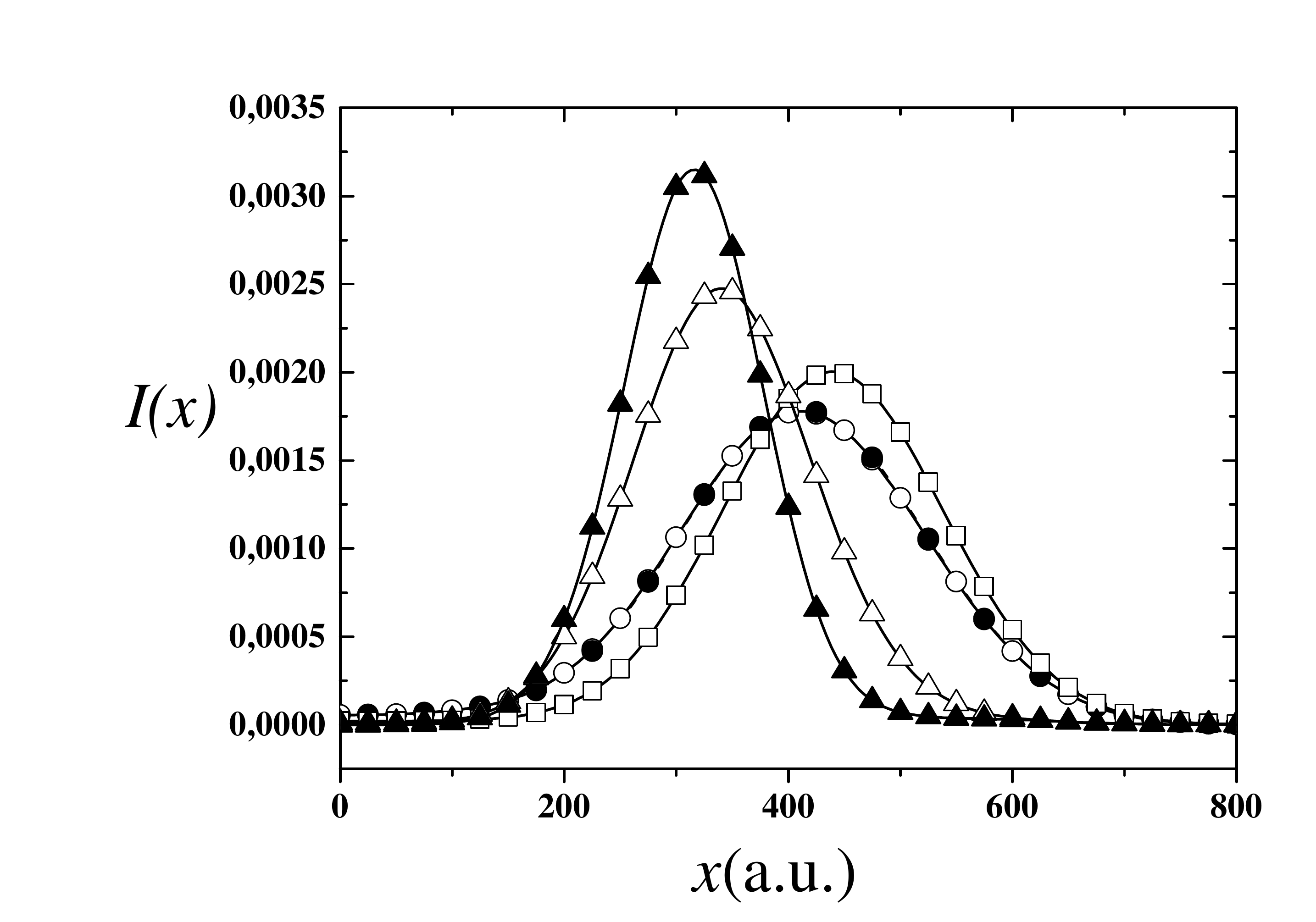}
\caption{The pair density distribution $I(x)$ of the singlet state as a function of $x$, where $x=x_1-x_2$ (see the text for details),  under the gaussian confinement potential for various values of $B$ in Tesla (T): $B=0$ (open circle), $B=2.35$ (solid circle), $B=20$ (open square), $B=40$ (open triangle) and $B=60$ (solid triangle),}
\label{I_GIB=0-60}
\end{figure}

\begin{figure}
\vspace*{-0.6 cm}
\hspace*{-1.2 cm}
\includegraphics[scale=0.37]{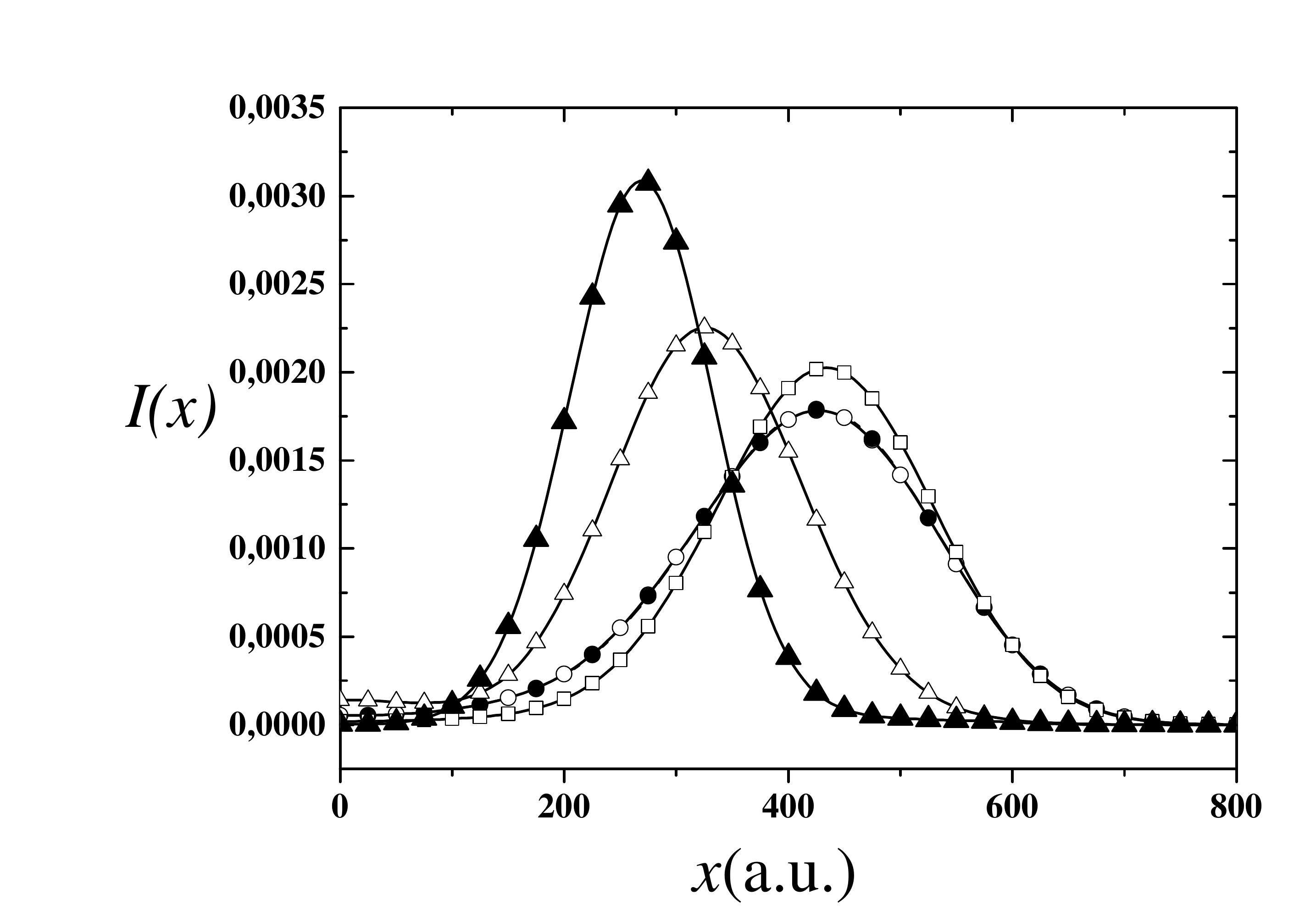}
\caption{The same as in Fig.~\ref{I_GIB=0-60} under the quartic confinement potential instead of the gaussian one.}
\label{I_QB=0-60}
\end{figure}

Now, we turn our attention to the average spatial distribution of the two electrons. As mentioned at the end of subsection \ref{I(x)}, the issue of double-occupation of a single dot by the two electrons is highly undesirable because the dots play the role of labeling the electron spins in two-qubit quantum-gate operation. In Fig.~\ref{I_GIB=0-60} and Fig.~\ref{I_QB=0-60} we display the graphs of the pair density function $I(x)$ as a function of $x$, where $x$ is the difference between the $x$-coordinates of electron $1$ and $2$, i.e. $x=x_1-x_2$. As $I(x)$ is symmetric in $x$ we have displayed it only for $x>0$. 

In Fig.~\ref{I_GIB=0-60} we show the behaviour of $I(x)$ with respect to the singlet state under the gaussian confinement potential for various values of $B$ up to $60$~T. For low magnetic field, $B=0$~T and $2.35$~T, the pair density distribution reveals the electrons at a distance smaller than the dots distance; the peak of $I(x)$ is approximately at $x=400$~a.u. while the dots distance corresponds typically to twice the value of $X_Q$, i.e. $\approx 480$~a.u.. By increasing the magnetic field strength, the pair density distribution shifts to right, towards larger values of $x$, which means that the electrons move away from each other. This displacement to right of $I(x)$ reaches its maximum around $B=20$~T, whose peak of the pair density distribution is at $x\approx 450$~a.u., which corresponds in good approximation to the electrons at different dots. As the strength of the magnetic field increases further, $B=40$~T and $60$~T, the electrons move closer together corresponding a displacement of $I(x)$ to left. Hence, we see that, according to the criterion of labeling the electron spins, the best range of magnetic field strength to work with is around $B=20$~T. On the other hand, at $B=20$~T the value of $J$ is quite low, being close to its minimum in the range of magnetic fields where $J$ does not approach zero. For this configuration of confinement potential one sees clearly a compromise relationship between the priorities involved: the two electrons well separated and exchange coupling values $J$ as high as possible. This compromise relationship leads us to see the behavior with the same values of magnetic field when the electrons are under the confinement of the quartic potential.

In Fig.~\ref{I_QB=0-60} it is shown the behaviour of $I(x)$ with respect to the singlet state under the gaussian confinement potential for various values of $B$ up to $60$~T. In general the behaviour is similar with that observed in Fig.~\ref{I_GIB=0-60}. However we can observe that the distribution $I(x)$ remains practically unchanged for magnetic field in the interval from $0$ to $20$~T with its peak at $x \approx 425$. For values greater than $20$~T the pair density distribution moves to left in the same fashion as in the gaussian confinement case. Thus, as long as $I(x)$ concerns, there is no difference whether one works with $B=0$~T or $B=20$~T. Now, if we look at Fig.~\ref{J versus B}, we see that, for the quartic potential case, the best region of magnetic field to work with is around $B \approx 0$~T. Therefore, in this case we do not have the issue of the compromise relationship mentioned above. Certainly this a consequence of $V_Q$ not being bounded from above. Therefore, these results give a clue of an important detail in designing a double QD: the size of the lateral side of the confinement potential is important and not just the internal barrier size.

Finally, notice that the width of $I(x)$ for larger field values is smaller than for low ones;  this is expected due to the stronger confinement caused by the higher value of $B$.

\section{Conclusion}
We have analysed the behaviour of a system consisted of two coupled quantum dots, each containing one excess electron, in the presence of a magnetic field. This system is known as two-electron double QD and is a good candidate to be the primary logic element of quantum computers. Our first and more evident objective was to see how the singlet and triplet energies of this system behave as a function of the magnetic field for typical confinement potential profiles that one founds in literature. Connected with that, another purpose was to see the influence of the confinement profile on some other quantities; we have focused on the exchange interaction $J$ and the electron pair density along the dot separation axis, $I(x)$. As a consequence of that, we have discussed how appropriate the two-electron double QD is robust as a basic tool for two-qubit quantum-gate operation depending on the range of magnetic field strength.

Basically what we have seen is that the energies of the singlet and triplet states have the same behaviour for both confinement potential considered, which is not a surprise. Their values diverge as the field strength $B$ increases, but remain relatively close in all range of $B$ analysed. On the other hand, when it comes to the exchange interaction $J$, which is the difference between the triplet and singlet energies, i.e., $E_T(B)-E_S(B)$, there is a range of $B$ where the values obtained with a confinement potential is quite different of that obtained with the other one. As in real situation the exact profile of the confinement is unknown, this results points that the range of magnetic field $25\textrm{~T} \lesssim B \lesssim 60\textrm{~T}$ is not proper because the values of $J$ are quite different  depending on the confinement profile. Besides, for small values of $B\; (B\lesssim 1\textrm{~T})$ one gets high values of $J$, which is desirable concerning the uncertainty issue involved in any experimental measure.  Finally, when one takes into consideration the one dot double-occupation by analysing the behaviour of the pair density function $I(x)$, one sees that, depending on the potential profile, one has or not a compromise relationship between the control and precision of the exchange coupling $J$ (the higher its value, the better to control and measure $J$) and the labeling capacity of the two electrons. Both characteristics are important for two-qubit quantum-gate operations. Thus, the importance of the deepness of the double QD well as a whole plays a more important role than might at first be expected.

It is worth mentioning that our choice of confinement potential involved the parabolic-quartic potential, which is only delimited below and, thus, does not support electron unbound states. On the other hand, the gaussian confinement potential profile constituted by two inverted gaussians contemplates the possibility of having unbound states from $\approx 640\; \textrm{meV}$. \\ \\

\bibliographystyle{iopart-num}
\bibliography{RefQDots5}

\end{document}